\documentclass[aps,prl,groupedaddress]{revtex4-2}

\usepackage{graphicx}
\usepackage{subcaption}
\usepackage{placeins}
\usepackage{siunitx}
\DeclareSIUnit{\pixel}{px}
\usepackage{lipsum}
\usepackage{soul}
\usepackage{float}
\usepackage{color}
\usepackage{xcolor}
\usepackage{amsmath}
\usepackage{cancel}
\usepackage{amsfonts}
\usepackage{ragged2e}
\def\({\left(}
\def\){\right)}
\def\[{\left[}
\def\]{\right]}

\begin{document}

\title{Butter on a hot pan: self-regulating dynamics of melt-lubricated sliding}

\author{Edoardo Bellincioni}
\email[Corresponding author: ]{\href{mailto:e.bellincioni@utwente.nl}{e.bellincioni@utwente.nl}}
\author{Simon Biermann}
\author{Jacco H. Snoeijer}
\author{Leen van Wijngaarden}
\author{Sander G. Huisman}
\email[Corresponding author: ]{\href{mailto:s.g.huisman@utwente.nl}{s.g.huisman@utwente.nl}}

\affiliation{Physics of Fluids Department and Max Planck Center for Complex Fluid Dynamics and J.M. Burgers Centre for Fluid Dynamics, University of Twente, P.O. Box 217, 7500AE Enschede, The Netherlands}

\date{\today}

\begin{abstract}
When solids melt while sliding down heated inclines, their motion is governed by a complex coupling between heat transfer, phase change, gravity and viscous dissipation. Despite relevance across a variety of domains, like kitchen physics, geophysics, tribology, and manufacturing, this coupled problem lacks understanding and quantitative experimental validation. Here we report experiments with ice and paraffin wax on a temperature-controlled ramp that achieve terminal velocities from \qty{0.01}{\meter\per\second} to \qty{2}{\meter\per\second} across wide parameter ranges. We develop a theoretical model that captures the self-regulating feedback between melt-layer thickness, sliding velocity, and heat transfer. Without any adjustable parameters, our model collapses all measurements, validating the fundamental mechanism and enabling predictions for analogous systems.
\end{abstract}

\maketitle
\begin{figure*}
    \centering
    \includegraphics[width=\linewidth]{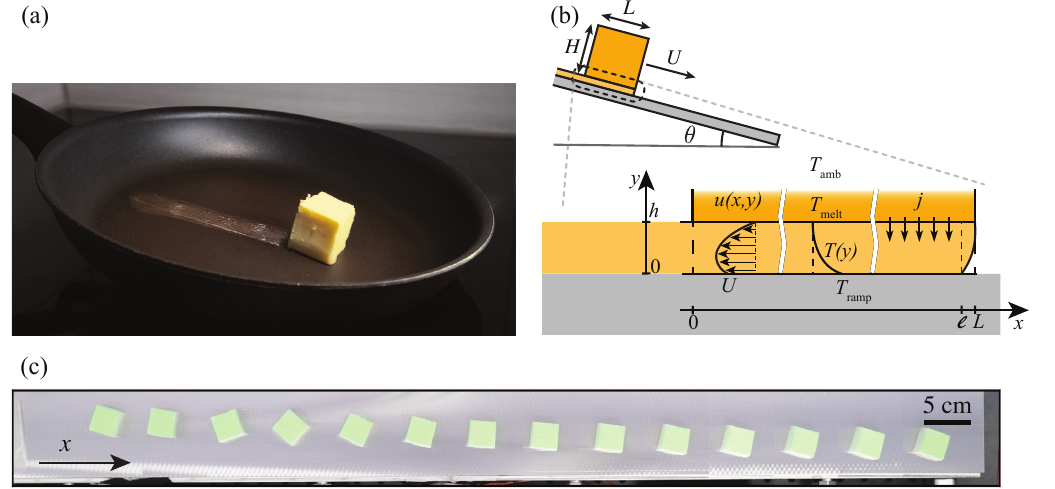}
    \caption{(a) A block of butter slides down an inclined, heated cooking pan due to the lubricating effect of the liquid melt layer beneath it. (b) Schematic of the problem, with the relevant physical quantities. A cubic block of solid with sides $L=H$ slides down a heated ramp, which is inclined at an angle $\theta$, with a constant velocity $U$. The enlargement is tilted to be aligned with the ramp and is in the frame of reference of the block.  $u(x,y)$ indicates the flow velocity; $T_\text{amb}$ is the ambient temperature; $T_\text{ramp}$ is the ramp temperature; $T_\text{melt}$ is the melting temperature; $j$ is the melt rate (with units of velocity). The symbol $\ell$ bounds the region $[\ell,L]$ where capillarity effects are dominant. (c) Superposed frames of a paraffin block. The block is moving from left to right. The time between the frames is \qty{0.5}{\second}, the angle of the ramp is \qty{2.5}{\degree} and the excess temperature is \qty{5}{\kelvin}. }
    \label{fig:pan}
\end{figure*}

It is a cook's common practice to spread a block of butter by having it slide controllably on an inclined, hot pan (figure \ref{fig:pan}a). When a solid fat slides down a hot pan, it demonstrates a deceptively complex coupling between heat transfer, phase change, and lubrication. The solid is sustained by its own melt, which generates the lubricating layer and thus must be continuously produced to maintain motion. Note that this is a self-regulating system: changes in sliding velocity alter transit time, modifying heat transfer and hence melt rate, which, in turn, adjusts layer thickness and viscous resistance --- feeding back to velocity. 

The ubiquity of this type of thermal-fluid-mechanical coupling is striking. Geophysical examples are lava flows \cite{griffiths_effects_1993,moyers-gonzalez_non-isothermal_2022}, where there is a liquid flow which solidifies at its interfaces, and glacier sliding \cite{Hallet_1979,Weertman_Birchfield_1983}, where immense pressures are the cause for the liquid--solid phase change. 
A similar mechanism is encountered in the Leidenfrost effect, where a vaporizing liquid (or sublimating solid) is levitated by its own vapour \cite{parrenin_dry_2021,dupeux_self-propelling_2013}, resembling the lubrication of an air hockey puck, which is sustained by an externally-imposed air flow from a porous plate \cite[ch. 5]{Leal_2007},\cite{petit1986sustentation,hinch_effect_1994,lemaitre_air_1990,wang_porous_2012}. 
Self-regulated lubrication is also commonplace in engineering applications such as lubricant bleeding or exudation \cite{lugt_review_2009,zhang_model_2021,camousseigt_oil-bleeding_2023}, where oil migrates out of the thickener network, forming a thin viscous film between the grease and the substrate, or in friction stir welding where molten material lubricates the motion and forms the weld itself \cite{kilic_comprehensive_2025}.
Lastly, in close-contact melting \cite{moallemi1986analysis}, a phase-change material is placed on top of a surface heated above its melting point, and the motion of the material is controlled externally. 
These problems share some similarities but also differ in important ways. In all the cases, a mass is sustained by a fluid film, but the film can either be externally provided or extracted from the mass itself. 

Despite the ubiquity of melt-lubricated motion across scales from micrometers to kilometers, the fundamental coupling between heat transfer, phase change, and viscous flow is challenging to study in controlled settings.
In this Letter, we show that many of these aspects can be studied with a surprisingly simple system, that is directly inspired by the sliding of melting butter on a hot pan. We report on idealised experiments performed with blocks of paraffin wax and ice melting down a temperature-controlled ramp, and we provide a theory that quantitatively describes the melt-lubricated sliding. 

\paragraph{Experiments}
We conducted experiments using blocks of ice and paraffin wax (the physical properties of the latter are in the End Matter) of a few centimetres in length on an \qty{1}{\meter}-long inclined stainless steel ramp, see top of figure \ref{fig:pan}b. To ensure an accurate and homogeneous temperature across the surface the ramp incorporates numerous internally distributed fluidic channels, through which a water--ethylene-glycol mixture is flowing. The temperature of the ramp is measured with a K-type thermocouple; the position of the blocks is tracked over time with a DSLR camera mounted above the ramp,  and the angle of inclination is measured with a digital level. See Supplemental Material \cite{supp} for further experimental details. 

\begin{figure*}
    \centering
    \includegraphics[width=\linewidth]{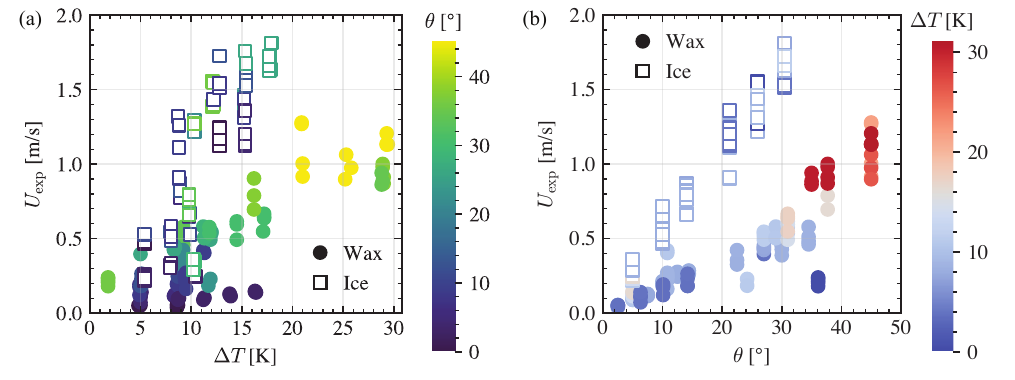}
    \caption{Terminal velocity of ice (hollow squares) and paraffin (disks) blocks sliding down an inclined heated ramp as a function of excess temperature $\Delta T=T_\text{ramp}-T_\text{melt}$ (a) and inclination angle $\theta$ (b). In both plots, the variable that is not in abscissa is shown as the markers' colour. As the blocks are at terminal velocity, the errors in the measured velocities are tiny (less than 1\%), and no error bars are shown. }
    \label{fig:rawData}
\end{figure*}

The experiments start with the gentle deposition of a solid block of ice or wax on the ramp. The ramp temperature $T_\text{ramp}$ is higher than than the melting temperature $T_\text{melt}$ of the sliding block. An example of an experiment is shown in figure \ref{fig:pan}c. After an initial transient, where heat diffuses into the solid and starts melting it, the thickness of the newly-created melt layer is enough to lubricate the friction between the block and the ramp, and the block will start sliding down. All the blocks reach terminal velocity before halfway down the ramp, hence we determine the terminal velocity by using a linear fit of the position for the last half. During the transient phase, some of the blocks are seen to rotate around the normal to the ramp. We do not observe any change on the dissipation dynamics nor on the heat transfer for repeated experiments. Figure \ref{fig:rawData} reports the measured velocity as a function of angle ($2^\circ\leq\theta\leq45^\circ$) and excess temperature (temperature of the ramp above the melting temperature, \qty{1}{\kelvin}~$\leq T_\text{ramp}-T_\text{melt}\leq$ \qty{31}{\kelvin}) for both wax and ice samples. From the plot, we can infer that both an increasing inclination and increasing excess temperature result in an increasing velocity, and that due to the different physical properties between the two investigated materials (primarily viscosity), ice reaches a higher velocity. Figure \ref{fig:rawData} reveals that neither excess temperature nor inclination alone determines the velocity --- simple power laws fail, owing to the complex interplay between heat transfer, melt rate, and viscous dissipation. 

\begin{figure}
    \centering
    \includegraphics[width=0.6\textwidth]{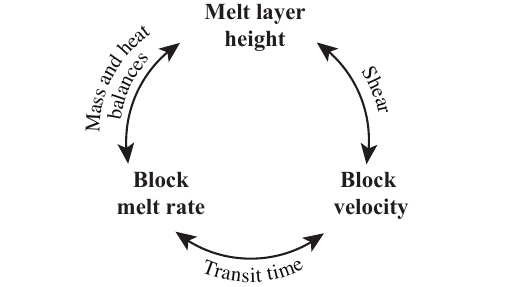}
    \caption{Schematic of the coupling between the effects of shear, transit time, and mass and heat balances. }
    \label{fig:coupledness}
\end{figure}

With the help of figure \ref{fig:coupledness}, we can give a qualitative explanation for the expected coupling between heat transfer, melt rate, and viscous dissipation, and for the stability of the system. Starting from the top of the figure, let us suppose that the thickness of the layer decreases; that would cause an increased shear, which would in turn decrease the velocity; a thinner layer and a slower block would mean a more effective heat conduction through the liquid, hence higher melt rate; due to mass conservation, a higher meltrate would increase the thickness of the layer; and the consequences would be as just described, but reversed. This self-regulatory, negative feedback ensures the stability of the system and the uniqueness of the triplets (height, melt rate, velocity). This also explains why, experimentally, the motion converges to a terminal velocity. 

\paragraph{Model}
We now develop a two-dimensional model that quantitatively captures the dynamics. The effect of neglecting the third dimension will be expanded later. Working in the frame of the sliding block (figure \ref{fig:pan}b), we apply lubrication theory \cite{kundu2016fluid,Leal_2007}, which describes the melt layer flow as superposed Couette and Poiseuille contributions. The region at the leading edge (extent $(L-\ell)=\mathcal{O}( h) \ll L$, see Supplementary Material \cite{supp}) is influenced by capillary effects due to the advancing contact line. We exclude this region from our model, as its integrated contribution to the total dissipation scales as $h/L\ll 1$, and focus on $x\in[0,\ell]$. The Couette component arises from the no-slip condition at the boundaries with a velocity difference $U$ and the Poiseuille component from pressure gradients needed to evacuate melt at the back:

\begin{equation}\label{eqn:flow}
u(x,y) = \frac{1}{2\mu}y(y-h)\frac{\partial p}{\partial x}+U\left(\frac{y}{h}-1\right)\ .
\end{equation}

Mass conservation allows to compute the pressure gradient. Since melt is assumed to generate uniformly at rate $j$ (units \unit{\meter\per\second}) but must escape at the trailing edge, the flux at position $x$ equals the integrated melt production from $\ell$ to $x$, under the condition of no mass flux at $x=\ell$. This is supported by experimental observations (see Supplemental Material \cite{supp}), and constitutes a significant deviation from the standard Leidenfrost modelling, which is due to the relatively slow production (and ejection) of melt compared to vapour. The mass conservation, combined with equation \ref{eqn:flow}, yields:
\begin{equation}\label{eqn:2}
    j(x-\ell)= \int_0^h u(x,y)dy = -\left(\frac{\partial p}{\partial x} \frac{h^3}{12\mu}+\frac{hU}{2}\right)\ .
\end{equation}
This expression can be integrated with the boundary condition that $p(x=0)=0$, to yield 
\begin{equation}\label{eqn:3}
    p(x)=-\frac{6\mu x}{h^3}\left[j(x-2\ell)+hU_0\right]\  .
\end{equation}

\begin{figure*}[!t]
    \centering
    \includegraphics[width=\linewidth]{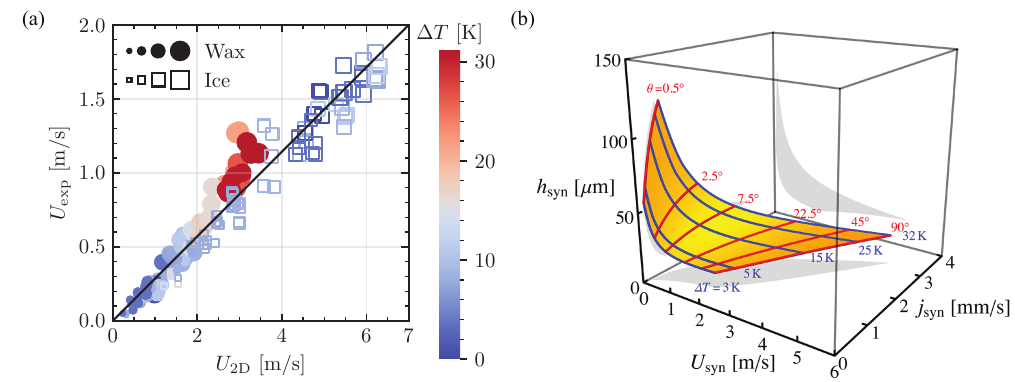}
    \caption{In panel (a): blocks' measured velocity $U_\text{exp}$ against the velocity from our analytical model $U_\text{2D}$; paraffin wax data are plotted as disks, while ice data are plotted as squares; the colour of the points indicate the excess temperature $\Delta T=T_\text{ramp}-T_\text{melt}$, while the size indicates the angle $\theta$, as in the legend (markers are \qty{2.5}{\degree}, \qty{7.5}{\degree}, \qty{22.5}{\degree}, \qty{45}{\degree}). The error bars on $U_\text{2D}$ were calculated with the bootstrapping method, but are not plotted as they never exceeded a marker's width. The rescaling does not involve any adjustable parameters; the solid line corresponds to $y=\frac{2}{7}x$.  In panel (b), for synthetic data with \qty{25}{\milli\meter} blocks of wax at angles in \qty{0.5}{\degree} $\leq \theta \leq$ \qty{90}{\degree}, and excess temperatures in \qty{3}{\kelvin} $\leq \Delta T \leq$ \qty{32}{\kelvin}, we calculate: the melt rate $j_\text{syn}$, the velocity $U_\text{syn}$, and the layer height $h_\text{syn}$, and we represent them as a 3D surface with iso-$\theta$ contours marked in red and iso-$\Delta T$ contours marked in blue. }
    \label{fig:collapsedData}
\end{figure*}

The apparent simplicity of the choice of boundary conditions of equations \ref{eqn:2} and \ref{eqn:3} at the leading edge hides interesting physics. In equation \ref{eqn:2}, we impose the flux to be zero at $x=\ell$. However, for a flat plate ($\theta=0$) one would expect the leakage to be equal on both sides, with the ratio between the two shifting with the inclination of the ramp. Yet, deriving the equations of motion allowing for leak at the leading edge reveals that the front leak becomes negative (inadmissible, as no material exists ahead of the sliding block) for an inclination angle of a fraction of a degree ($\theta_\text{crit}=\mathcal{O}(\ang{;;10})$).

A global force balance closes the mechanical problem. Normal to the ramp, pressure supports the block's weight 
\begin{equation}\label{eqn:normal_momentum}
    mg\cos{(\theta)}= W\int_0^\ell p(x)dx=\frac{\mu W\ell^2}{h^3}\left(4j\ell-3hU\right)\  ,
\end{equation}
with $W$ the width of the block into the page. Tangentially, gravity is balanced by viscous shear  stress \begin{equation}\label{eqn:tangential_momentum}
    mg\sin{(\theta)}= W\mu\int_0^\ell\frac{\partial u}{\partial y}\bigg\rvert_{y=h} dx =\frac{\mu W\ell}{h^2}\left(3j\ell -2hU\right)\  .
\end{equation}

Throughout this derivation, we have assumed that the mass of the block does not change during the experiment. This assumption is based on the large separation of scale between the typical speed of the experiment $\approx$ \qty{1}{\meter\per\second} and the typical interface velocity $\approx$ \qty{1}{\milli\meter\per\second}, estimated from the typical duration of the experiment ($\approx$ \qty{1}{\second}) and a typical height decrease ($\approx$ \qty{1}{\milli\meter}, which is $\leq5\%$ of the block thickness).

The thermal boundary condition couples mechanics to phase change. At the solid--liquid interface, the Stefan condition \cite{carslaw_conduction_1959} equates heat flux to melt rate
\begin{equation}\label{eqn:heat_transfer}
    k_l \frac{\partial T}{\partial y}\bigg|_{y=h^-}= j \mathcal{L}\rho_l\ ,
\end{equation}
with $k_l$ the thermal conductivity in the liquid, $\mathcal{L}$ the latent heat, and $\frac{\partial T}{\partial y}$ the local temperature gradient at the top liquid-solid interface $y=h^-$. For the latter, the steady state expression $\frac{T_\text{ramp}-T_\text{melt}}{h}$ provides a good estimate, which can be further refined considering the finite-time heat diffusion during the transit time $t_\text{tr}=L/U$, although for our experiments this correction is always tiny. A detailed derivation of the correction is presented in the Supplemental Material \cite{supp}. 
Since $\frac{T_\text{ramp}-T_\text{melt}}{h}\gg\frac{T_\text{melt}-T_\text{amb}}{H}$ in our experiments, the heat transfer in the solid part of the block is negligible. 

\paragraph{Collapse of data}
Equations \ref{eqn:normal_momentum}--\ref{eqn:heat_transfer} constitute the governing equations for our problem, and are functions of the variables $U$, $j$, and $h$, with $T_\text{ramp}$ and $\theta$ being the control parameters. Algebraic steps reduce the system to fourth-order equations for the unknowns, which are symbolically solved. The scattered experimental data of figure \ref{fig:rawData} can now be collapsed.
Figure \ref{fig:collapsedData}a compares the experimental velocity with the theoretical predictions, for both paraffin and ice. There is good collapse of the experimental data, for both materials, and across all the investigated ranges of ramp inclination angle and excess temperature. The collapse occurs for both ice and paraffin, which have substantially different wettability properties on stainless steel: the water film rapidly dewets to form partially wetting drops, while the wax is appears to be completely wetting and leaves a persistent film. This strengthens the validity of our choice of neglecting capillarity effects, and highlights the generality of our model for different surface treatments. We remark that surface tension effects are expected to become dominant over gravity at very small angles of inclinations, but such regime lies out of the scope of this work. Compared to the data, our two-dimensional model overestimates the experimental velocities by a factor $\approx$\num{3.5}. Such a factor can be rationalized by the 2D approximation of the pressure profile. In the direction into the page direction, the pressure has as boundary conditions $p=0$ at the two ends, whereas we assume a flat pressure profile: this suggests an overestimation of the mean pressure and an underestimation of the pressure gradient, both of which lead to decreased viscous dissipation. 

To investigate the predictive power of our model, in figure \ref{fig:collapsedData}b, we plot the 3D surface defined by the triplets $(U,j,h)$ for wax, calculated with synthetic $(\Delta T, \theta)$ couples in the ranges [\qty{3}{\kelvin}, \qty{32}{\kelvin}] and [\qty{0.5}{\degree}, \qty{90}{\degree}], for $\Delta T$ and $\theta$, respectively. The key observations are that $(U, h, j)$ lie on a uniquely-defined surface of solutions, and that the iso-$\Delta T$ or iso-$\theta$ contour lines do not intersect. Hence $(U, h, j)$ form unique triplets: for given external conditions $(\Delta T, \theta)$, the system admits only one equilibrium state where all balances are satisfied. This uniqueness confirms the self-regulating mechanism of figure \ref{fig:coupledness}.

Our model predicts layer thicknesses of \num{10}--\num{100}\unit{\micro\meter} and melt rates of $\mathcal{O}$(\qty{1}{\milli\meter\per\second}) depending on conditions, but precise measurements remain challenging. For the former, the modelled values are consistent with our experimental observations, see Supplemental Material. For the latter, we can estimate from experiments that a block's height decreases $\mathcal{O}$(\qty{1}{\milli\meter}) during a time that is  $\mathcal{O}$(\qty{1}{\second}), providing a matching estimate.

In summary, we investigated the problem of self-lubricated motion of a melting object, which is intrinsically characterised by the coupling between thermal, mechanical, and fluid dynamics. The simplicity of the problem, and the consequent experimental accessibility, allowed us to thoroughly control the otherwise intricate three-way coupling. We find good collapse of the experimental results on the 2D analytical model, across 100-fold velocity variations and different materials. 

In favour of simplicity, we neglected some mechanisms which could be included in a more refined version of the model. For example: 
we neglect leakage from the sides of the block; 
we assume a flat solid-melt interface, with $j$ not depending on position, and neglect both bottom curvature and tilt;
we neglect the alteration to the temperature profile from the vertical convective motion induced by the melting; 
we do not consider the viscosity changes within the vertical temperature profile;
and we do not consider the torque that is exerted on the block by the shearing liquid beneath it. 
Despite these assumptions, the non-trivial collapse of the experimental data of figure \ref{fig:collapsedData}, shows that our analytical model captures the essential physics of the problem. 

Concluding, beyond clarifying a commonplace phenomenon, like butter on a pan, this work establishes a quantitative model system for melt-lubricated dynamics. Unlike similar systems, the current study is experimentally accessible and reproducible, making it an ideal testbed for validating theories applicable to those inaccessible. The parameter-free collapse we demonstrate suggests our framework captures essential physics applicable to geophysical flows, tribological systems, and manufacturing processes where direct validation is impractical. Further work could proceed in this direction.

\begin{acknowledgments}
\paragraph{Acknowledgements}
We thank Christian Diddens for valuable insights, Gert-Wim Bruggert, Martin Bos, and Thomas Zijlstra for technical support, Coen Verschuur for the viscosity measurements, and Hyunjong Lee for the TPS measurements. E.B. wishes to thank Stefaniia Lozenko for the help in capturing the photo in figure \ref{fig:pan}a. 

This work was financially supported by the European Union (ERC, MeltDyn, 101040254).
\end{acknowledgments}
\newpage
\appendix
\section{End Matter }\label{sec:endmatter}
\begin{table}[h]
\caption{\label{tab:physProperties}Physical properties of the paraffin wax used in the experiments. The properties were measured as follows: densities were measured with an Archimedes setup (at \qty{20}{\celsius} for the solid and \qty{59}{\celsius} $\pm$ \qty{1}{\kelvin} for the liquid); latent heat with a Dewar vessel; thermal diffusivity and thermal conductivity with Thermtest Transient Plane Source instrument; and viscosity with an Anton Paar viscometer. }
\renewcommand{\arraystretch}{1.2}
\begin{ruledtabular}
\begin{tabular}{lccc}
\textit{property}&\textit{symbol}&\textit{value}&\textit{unit}\\ \hline
density of solid &$\rho_s$&\num{842} $\pm $ 1\%&\unit{\kilo\gram\per\meter\cubed}\\
density of liquid&$\rho_l$&\num{764} $\pm $ 5\%&\unit{\kilo\gram\per\meter\cubed}\\
latent heat&$\mathcal{L}$&\num{180e3} $\pm $ 5\%&\unit{\joule\per\kilo\gram}\\
thermal conductivity&$k$&\num{0.291} $\pm $ 4\%&\unit{\watt\per\meter\per\kelvin}\\
thermal diffusivity&$\kappa$&\num{0.094e-6} $\pm$ 4\%&\unit{\meter\squared\per\second}\\
specific heat capacity&$c_p$&$\frac{k}{\rho_l \kappa}= $ \num{4.05e3} &\unit{\joule\per\kilo\gram\per\kelvin}\\
melting temperature&$T_m$&44 $\pm $ \qty{1}{\kelvin}&\unit{\celsius}\\ \hline 
viscosity fit &\multicolumn{2}{c}{$aT^3+bT^2+cT+d$}&\unit{\meter\squared\per\second}\\
\qty{45}{\celsius}$\leq T\leq$\qty{75}{\celsius}&$a$&\num{-1.69e-5}&\unit{\meter\squared\per\second\per\kelvin\cubed}\\
&$b$&\num{4.36e-3}&\unit{\meter\squared\per\second\per\kelvin\squared}\\
&$c$&\num{- 4.15e-1}&\unit{\meter\squared\per\second\per\kelvin}\\
&$d$&\num{16.40}&\unit{\meter\squared\per\second}\\
\end{tabular}
\end{ruledtabular}
\end{table}
%

\widetext
\clearpage
\begin{center}
\textbf{\large Supplemental Material of:\\
Butter on a hot pan: self-regulating dynamics of melt-lubricated sliding}
\end{center}
\setcounter{equation}{0}
\setcounter{figure}{0}
\setcounter{table}{0}
\setcounter{page}{1}
\makeatletter
\renewcommand{\theequation}{S\arabic{equation}}
\renewcommand{\thefigure}{S\arabic{figure}}
\renewcommand{\bibnumfmt}[1]{[S#1]}
\renewcommand{\citenumfont}[1]{S#1}
\makeatother 

\section{Experimental Setup}
\begin{figure}[H]
    \centering
    
    \begin{subfigure}{0.48\linewidth}
        \centering
        \includegraphics[width=\linewidth]{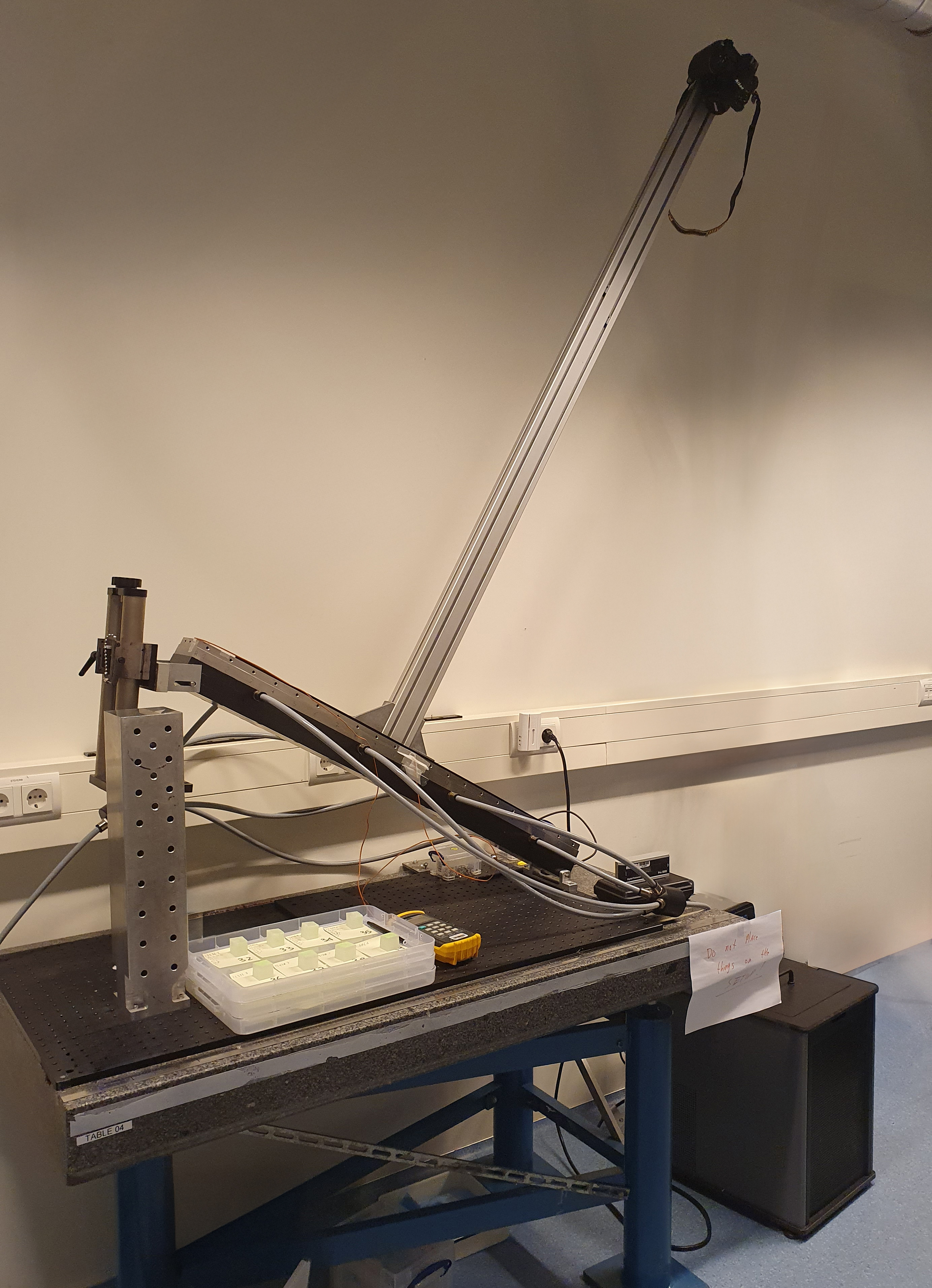}
        \caption{Photo of the experimental setup. The ramp is seen inclined. The DSLR camera is fixed on a bar mounted in the middle of the ramp. The heating unit is visible on the right, connected with the ingress and egress of the ramp via the grey tubes. Paraffin wax samples are visible on the table, together with the multimeter for the K-type thermocouple.}
        \label{fig:setup}
    \end{subfigure}
    \hfill
    \begin{subfigure}{0.48\linewidth}
        \centering
        \raisebox{0pt}[\dimexpr\height/2][\dimexpr\height/2]{%
            \includegraphics[width=\linewidth]{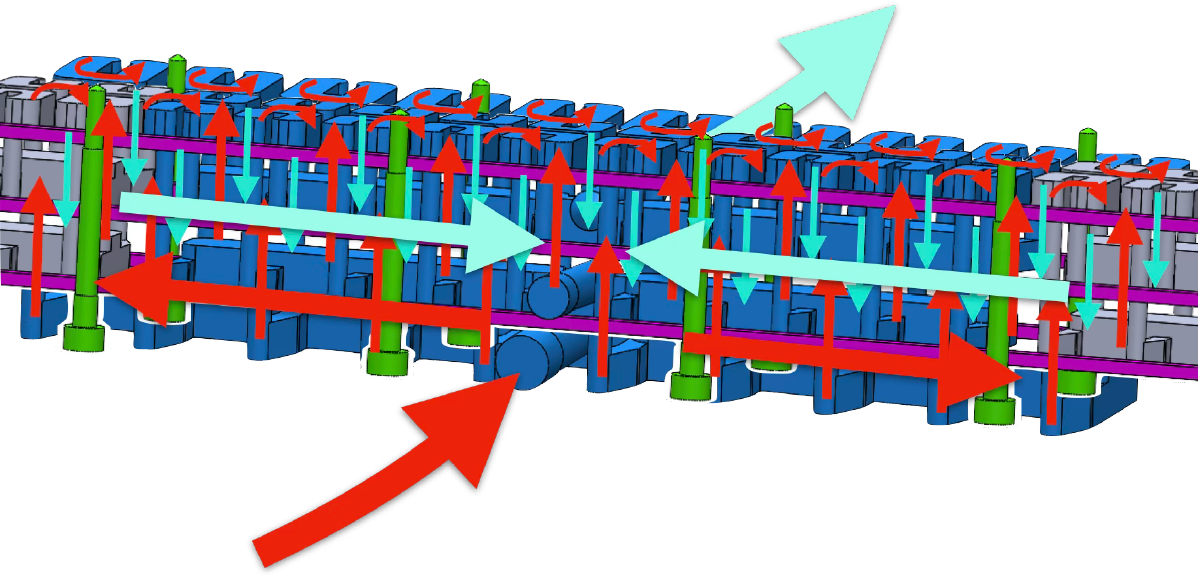}%
        }
        \caption{Coloured rendering of the inside of the ramp. Hot fluid (red) enters from the front, gets distributed in different heat-exchange chambers, then gets re-collected and exits from the back (cyan).}
        \label{fig:rampchannels}
    \end{subfigure}
    \caption{(a) Photograph of the experimental setup and (b) internal rendering of the ramp.}
    \label{fig:setup_combined}
\end{figure}

A photo of the setup and a rendering of the inside of the ramp are provided in figure \ref{fig:setup_combined}. When doing experiments with ice, the setup was moved inside a \qty{9}{\meter\cubed} temperature-controlled room, with a measured temperature of \qty{0}{\celsius} $\pm$ \qty{0.5}{\kelvin}. 
The blocks were created by melting the material in moulds and then cutting them to shape. Their length was then measured with a manual caliper (accuracy \qty{0.05}{\milli\meter}) and their weight with a laboratory scale (Sartorius QUINTIX5102-1S, accuracy \qty{0.01}{\gram}).

We used a PolyScience PD15R-30-A12E refrigerating and heating recirculating bath, with a temperature stability of $\pm$\qty{0.005}{\kelvin} in the range \qty{-30}{\celsius} to +\qty{200}{\celsius}.
We used a Nikon D850 camera (\qty{45.7}{\mega\pixel}, pixel size \qty{4.35}{\micro\meter}) with a Nikon AF Nikkor 50 mm f/1.8D lens (resulting in a resolution of approximately  \qty{30}{\micro\meter\per\pixel}), recording videos in 1080p at 60fps or 4K at 30fps, depending on the speed of the blocks. The top of the samples (both wax and ice) was marked green to help with tracking. 
The digital level was a Mitutoyo Pro 3600, with an accuracy of \qty{0.05}{\degree} in the range \qty{0}{\degree} to \qty{10}{\degree} and of \qty{0.2}{\degree} in the range \qty{10}{\degree} to \qty{40}{\degree}.
\section{Unsteady heat transfer}
Carslaw and Jaeger \cite[pp. 99--100]{carslaw_conduction_1959} reports an analytical expression for the temperature evolution in a bounded one-dimensional region, with a given initial temperature and given Dirichlet boundary conditions at both ends. The solution is composed of the sum between a steady term ($T_\text{st}$) and a time-dependent term ($T_\text{t.d.}$). With reference to the sketch in the right panel of figure 3 of the paper, we consider that the initial temperature of the liquid is that of the melt, and, during the time of transit of the block above a point, heat is allowed to flow through the liquid, increasing its temperature. Our initial and boundary conditions are of $T(y=0,t)=T_\text{ramp}$, $T(y=h,t)=T_\text{melt}$, $T(y,t=0)=T_\text{melt}$. The expressions for the steady and time-dependent temperature terms are
\begin{align}
    T_\text{st}(y)&= \frac{T_\text{melt}-T_\text{ramp}}{h}y+T_\text{ramp}\\
    T_\text{t.d.}(y,t)&= \sum_{n=1}^{\infty}a_n\sin\left(\frac{n\pi y}{h}\right)\exp\left(-\frac{\kappa n^2\pi^2t}{h^2}\right)
\end{align}
with $\kappa$ the thermal diffusivity, and

\begin{align}
    a_n&=\frac{2}{h}\int_0^h\left(T_\text{melt}-T_\text{ramp}\right)\left(1-\frac{y'}{h}\right)\sin\left(\frac{n\pi y'}{h}\right)dy'=\\
    &= \frac{2\left(T_\text{ramp}-T_\text{melt}\right)\(n\pi-\cancel{\sin{(n\pi)}}\)}{n^2\pi^2}=\frac{2\left(T_\text{melt}-T_\text{ramp}\right)}{n\pi}\ ,
\end{align}
where the sine term cancels as $n\in\mathbb{N}$.

We note that time appears in a non-dimensional form which can be identified as the Fourier number $\text{Fo}$,
\begin{equation}
    \text{Fo}=\pi^2\frac{\kappa t}{h^2} .
\end{equation}
For the Stefan condition, we need the temperature gradient in the liquid, evaluated at the top solid-liquid interface, $y=h^-$. The contribution of the steady term is trivial, while the $y$-derivative of the time-dependent term reads
\begin{equation}
    \frac{\partial T_\text{t.d.}}{\partial y}=\frac{2\left(T_\text{melt}-T_\text{ramp}\right)}{h}\sum_{n=1}^{\infty}\cos\left(\frac{n\pi y}{h}\right)\exp\left(-n^2\text{Fo}\right) \ ,
\end{equation}
which, evaluated at $y=h^-$ is 
\begin{align}
    \frac{\partial T_\text{t.d.}}{\partial y}\Bigg |_{y=h^-}&=\frac{2\left(T_\text{melt}-T_\text{ramp}\right)}{h}\sum_{n=1}^{\infty}(-1)^n\exp\left(-n^2\text{Fo}\right)=\\
    &=\frac{\left(T_\text{melt}-T_\text{ramp}\right)\left[\vartheta_4(0,e^{-\text{Fo}})-1\right]}{h}\ ,
\end{align}
with $\vartheta_4(u,q)=1+2\sum_{n=1}^\infty(-1)^nq^{n^2}\cos(2nu)$ an elliptic theta function. 

We calculate the mean temperature gradient as the mean of the time-dependent temperature gradient over the transit time $t_\text{tr}=L/U$, plus the steady temperature gradient, hence
\begin{align}
    \bigg\langle\frac{\partial T}{\partial y}\bigg\rangle_\text{mean}= \bigg\langle \frac{\partial T_\text{st}}{\partial y} + \frac{\partial T_\text{t.d.}}{\partial y}\bigg\rangle &= \frac{T_\text{melt}-T_\text{ramp}}{h}\left(1+\frac{1}{t_\text{tr}}\int_0^{t_\text{tr}}\left(\vartheta_4(0,e^{-\text{Fo}(t)})-1\right)dt\right) \\
    &=\frac{T_\text{melt}-T_\text{ramp}}{h}\left(\frac{1}{t_\text{tr}}\int_0^{t_\text{tr}}\vartheta_4(0,e^{-\text{Fo}(t)})dt\right)\ .
\end{align}
\section{Observation of layer thickness}
Hereafter an expanded view of a paraffin wax block sliding towards the camera. The wax is seen above the ramp. The inclination angle was \qty{7.5}{\degree} and the ramp had an excess temperature of \qty{8.3}{\kelvin}. 
\vspace{4mm}

\centering{
\includegraphics[width=.5\textwidth]{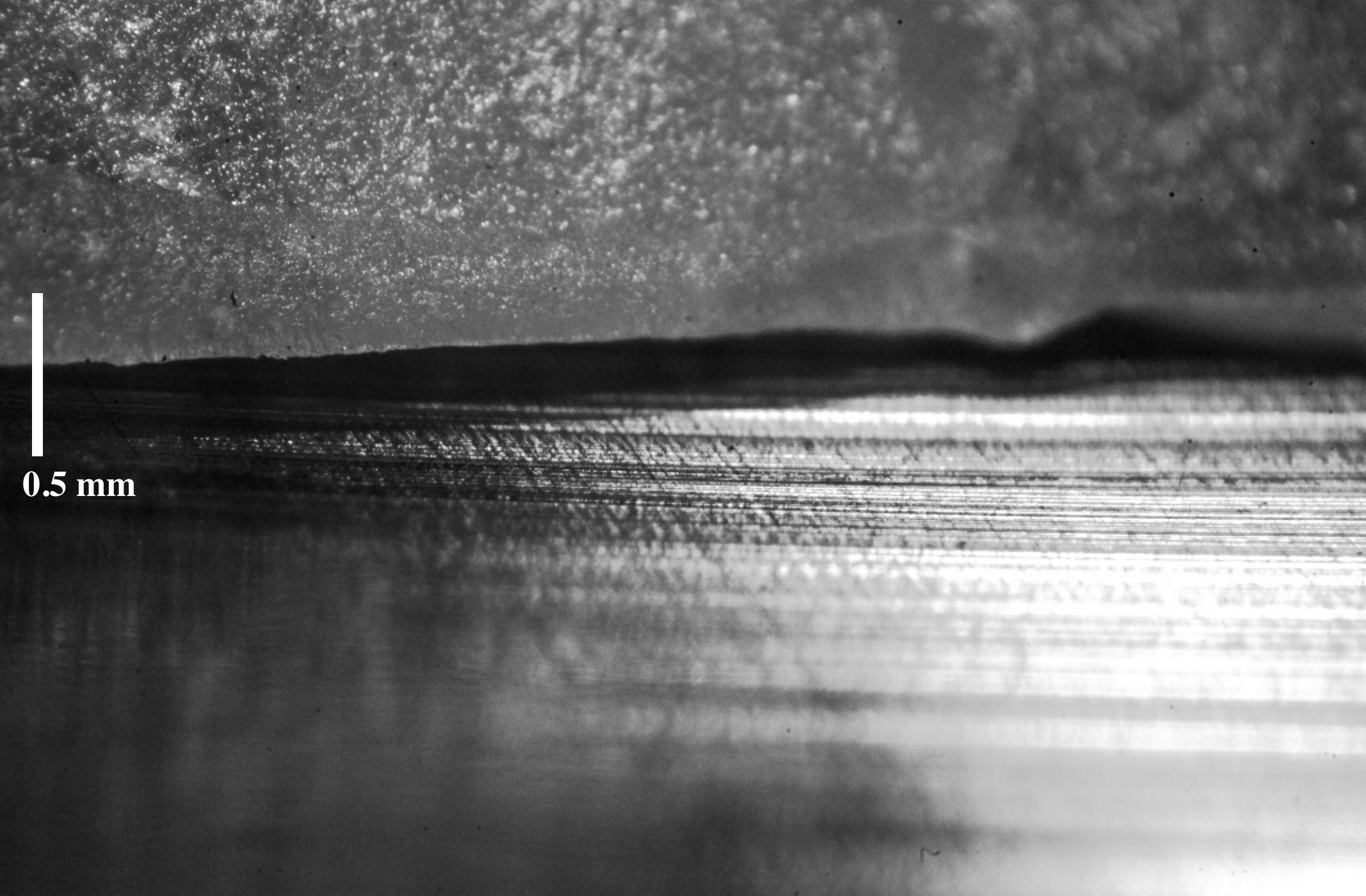}
}
\vspace{4mm}

\justifying
From the image, the distance of the block to the ramp can be estimated in the order of hundreds of micrometers. We also remark that the sample's surface imperfections are of the same order, hence this can only pose an upper bound for the thickness of the melt layer. 

\section{Observation of capillary region}
We performed an experiment on a heated, inclined,  \qty{7.3}{\centi\meter}-diameter sapphire glass plate, imaging the bottom of a wax sample with a high-speed camera. The temperature of the plate was controlled with a PID and set to an excess temperature of \qty{1}{\kelvin}. In the image below, where the wax is moving to the right, one can see from the right to the left: the leading side of the wax (the camera is oriented vertically, but the block slides down inclined), the contact line between the melt and the wax, and the contact line between the melt and the sapphire glass. We estimate a capillary region length in the order of hundreds of micrometers, and note that the properties and dimensions of the substrate differ from the ones of our experiment.
\vspace{4mm}

\centering{
\includegraphics[width=.3\textwidth]{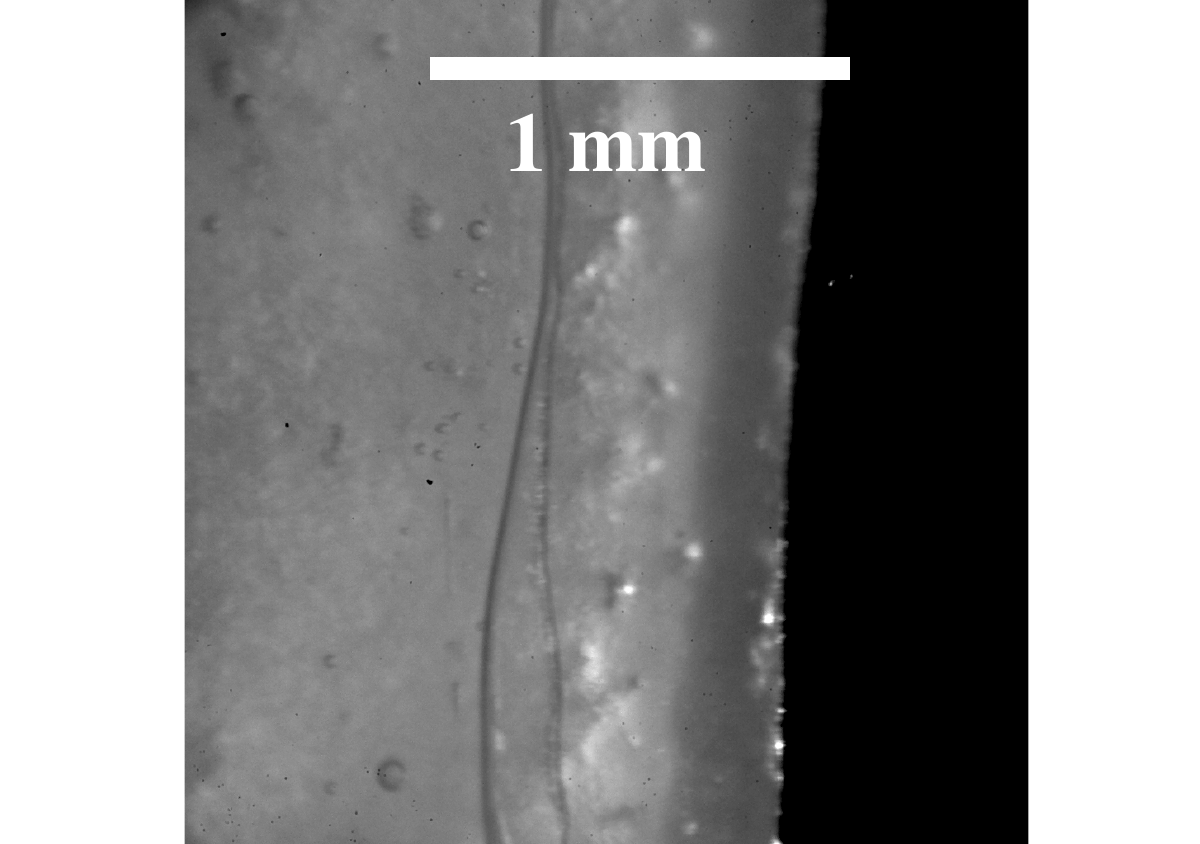}
}

\end{document}